\documentclass{moriond}

\usepackage{amsmath}
\usepackage{amsfonts}
\usepackage{amssymb}
\usepackage{xcolor}

\def\be{\begin{equation}}
\def\ee{\end{equation}}
\def\bea{\begin{eqnarray}}
\def\eea{\end{eqnarray}}

\newcommand{\LCDM}{$\Lambda$CDM}
\newcommand{\Photo}{}

\begin{document}
\vspace*{4cm}
\title{DO ANTHROPIC ARGUMENTS REALLY WORK?}

\author{ D. SORINI }

\address{Institute for Astronomy, University of Edinburgh, Royal Observatory, \\Edinburgh EH9 3HJ, United Kingdom}
\maketitle\abstracts{
The anthropic explanation for the peculiarly small observed value of the cosmological constant $\Lambda_{\rm obs}$ argues that this value promotes the formation of stars, planets, and ultimately of observers such as ourselves. I show that a recent analytic model of cosmic star formation predicts that although $\Lambda_{\rm obs}$ maximises the overall efficiency of star formation in the universe, the probability of generating observers peaks at $\sim400-500 \, \Lambda_{\rm obs}$. These preliminary results suggest that an immediate connection between star formation efficiency and observers' generation is not straightforward, and highlight the subtleties involved with the application of anthropic reasoning.
}

\section{Fundamental problems of the \LCDM\ paradigm}

Although the \LCDM\ paradigm has been remarkably successful at reproducing a plethora of observations, it is affected by fundamental and unsolved theoretical issues. Perhaps the biggest open question is the physical meaning of the cosmological constant $\Lambda$. The dominant view is that it is a manifestation of the energy of the quantum vacuum, but in this case the observed value $\Lambda_{\rm obs}$ is $\sim 120$ orders of magnitude smaller than the theoretical expectation. [~\cite{Barnes_rev}] There is still no consensus on a satisfactory explanation for the observed value of $\Lambda$ from first principles.

Alternatively, one can adopt an anthropic approach, positing that we observe such peculiarly small but non-null value of $\Lambda$ as a consequence of a selection effect: we are more likely to measure values that promote the generation of observers such as ourselves. As S. Weinberg noted in 1987, [~\cite{Weinberg_1987}] if $\Lambda$ were much larger than what we observe, then the accelerating expansion of the universe would occur at such early times that it would effectively prevent the formation of galaxies, stars, planets, and ultimately life (see also, e.g.,  [~\cite{Efstathiou_1995},~\cite{Peacock_2007}]). Translating this argument to the Multiverse scenario predicted by some models of stochastic inflation, only the universes with a suitably small $\Lambda$ would be populated with observers. But anthropic reasoning does not in itself require the physical existence of the Multiverse: one could still ask what values of $\Lambda$ would promote observers' generation among a hypothetical ensemble of possible universes. 

Making the minimal assumption that stars are necessary for producing observers, one can effectively consider star formation as a proxy for observers' generation. To test whether the anthropic approach can indeed explain the observed value of $\Lambda$, it is then necessary to understand how $\Lambda$ impacts the total efficiency of star formation over the entire history of the universe. Recently, this question was addressed by running suites of cosmological hydrodynamic simulations beyond redshift $z=0$, [~\cite{Salcido_2018}, \cite{BK_2021}] where $\Lambda$ was increased up to $300 \, \Lambda_{\rm obs}$. [~\cite{Barnes_2018}] Although impressive, such simulations still face some important limitations. The parameter space is restricted to a few values of $\Lambda$ due to the heavy computational cost. Furthermore, the latest cosmic time probed by simulations so far is $\sim 100 \,\rm Gyr$, but in principle one cannot exclude that there may be an important contribution to the global star formation history at even later times. [~\cite{Adams}, \cite{SP21}]. On the other hand, even though analytical models of cosmic star formation inevitably need to make some simplifying assumptions, they are not subject to the aforementioned restrictions. Therefore, they represent an attractive complementary approach for testing anthropic arguments. 

Using the recent analytic model for cosmic star formation given by Sorini \& Peacock (2021; hereafter, SP21), [~\cite{SP21}] I computed the total efficiency of star formation over the entire history of the universe, for a wide range of $\Lambda$. The preliminary results show that the efficiency is maximised when $\Lambda \approx \Lambda_{\rm obs}$. However, assuming a flat prior on $\Lambda$, the peak of observers' generation occurs in models with $\Lambda \sim 400-500 \, \Lambda_{\rm obs}$. This shows that maximum stellar efficiency does not automatically correspond to maximum observers' generation, highlighting the subtleties involved with invoking anthropic reasoning.

\section{Modelling past and future star formation in non-standard \LCDM\ models}

The SP21 analytical model of cosmic star formation allows a prediction of the star formation history in a flat \LCDM\ model with any $\Lambda$, and for arbitrarily large cosmic times ($t \rightarrow \infty$). It is based on an extension of the formalism introduced by Hernquist \& Springel (2003), [~\cite{HS03}] whereby the cosmic star formation rate density (CSFRD) is obtained as:
\begin{equation}
\label{eq:CSFRD}
	\dot{\rho}_{*} (z) = \bar{\rho}_0 \int s(M, \, z) \, \frac{dF(M, \, z)}{d\ln M} \, d\ln M ,
\end{equation}
where $\bar{\rho}_0$ is the comoving mean matter density of the universe, and $dF/d\ln M$ is the halo multiplicity function, with $F(M, \, z)$ being the collapse fraction in haloes with total mass $<M$. In the equation above, $s(M, \, z)$ is the star formation rate (SFR) within haloes of a fixed mass $M$, normalised by the total halo mass. Within the Hernquist-Springel formalism, $s(M, \, z)$ is set by whichever time scale is the shortest: an internal gas consumption time scale at high redshift, and the gas cooling time at low redshift. [~\cite{HS03}]

SP21 extended this formalism in two main aspects. First, they included a model for supernovae-driven winds that can alter the baryon mass fraction within haloes. Second, they explicitly considered the impact of the collapse time of haloes on the gas cooling time scale. These extensions result in an improved match to observations of the CSFRD, [~\cite{MD14}] which are reproduced within a factor of two over the redshift range $0<z<10$. More importantly, the model predicts a convergent time-integral of the CSFRD even for $t \rightarrow \infty$, and is therefore particularly well suited for testing anthropic arguments.

I thus adopt the SP21 model to compute the CSFRD for flat \LCDM\ models, with $10^{-4} < \Lambda / \Lambda_{\rm obs} < 10^4$. Modifying the cosmological constant clearly changes $\Omega_{\rm m}=1-\Omega_{\Lambda}$, and hence the evolution of the Hubble constant $H(z)$. Subsequently, both the mean matter density and the halo multiplicity function in Eq.~\ref{eq:CSFRD} are affected. Notably, altering $\Lambda$ also impacts the normalised SFR $s(M, \, z)$. This happens because the amount of gas that undergoes cooling is determined by following the expansion of a cooling front from the centre of the halo outwards. [~\cite{HS03}] The extent of the cooling front depends on the virial radius, whose redshift evolution is in turn set by $H(z)$. As such, the cooling rate is sensitive to the cosmological model via the Hubble constant.

The total stellar mass that will ever be produced in a unit comoving volume $\rho_{*\, \rm tot}$ can be straightforwardly obtained by integrating the CSFRD in Eq.~\ref{eq:CSFRD} from some initial time to $t\rightarrow \infty$. Equivalently, in terms of redshift:
\begin{equation}
    \rho_{*\, \rm tot} = \int ^{\rm z_{\rm in}} _{-1} \frac{\dot{\rho}_{*} (z) }{(1+z) H(z)} dz .
\end{equation}
For $z_{\rm in} \gtrsim 20$ the resulting $\rho_{*\, \rm tot}$ is already well converged. Dividing $\rho_{*\, \rm tot}$ by the comoving baryonic mass density of the universe $\bar \rho_{\rm b}$, one obtains the total stellar efficiency of that universe. 

\begin{figure}
    \centering
    \includegraphics[width=0.5\textwidth]{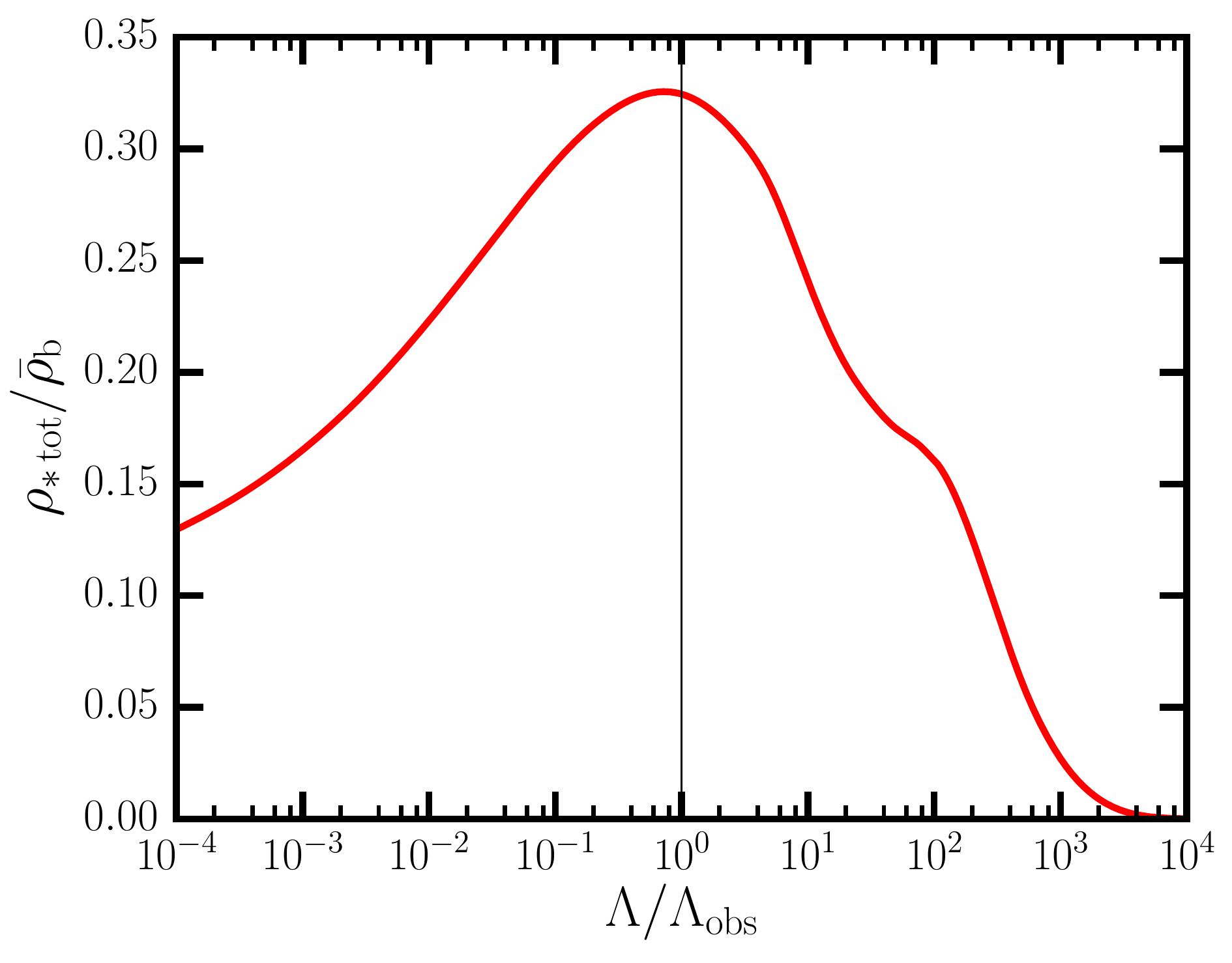}
    \caption{Fraction of baryonic mass density converted into stars over the entire history of the universe, as a function of the cosmological constant $\Lambda$, predicted by the Sorini \& Peacock (2021) model of cosmic star formation. [~\protect\cite{SP21}] Such global stellar efficiency peaks around the  observed value of $\Lambda$.}
    \label{fig:eff}
\end{figure}

Such total stellar efficiency is shown in Fig.~\ref{fig:eff}, as a function of $\Lambda$. Quite curiously, the maximum efficiency is achieved around the observed value of $\Lambda$, which is marked in the plot with a vertical black line. However, it is noteworthy that the efficiency is not suddenly suppressed for $\Lambda>\Lambda_{\rm obs}$. Even for $\Lambda\sim 100 \, \Lambda_{\rm obs}$, around $\sim 15\%$ of the baryonic mass in the universe is still converted into stars. To understand the implications of this result for anthropic arguments, we need to convert the stellar efficiency in Fig.~\ref{fig:eff} into a probability for the generation of observers in an ensemble of universes with different $\Lambda$.

\section{Implications for anthropic arguments}

To associate a probability of observers' generation to universes with different $\Lambda$, it is first necessary to assume a prior on $\Lambda$. I assume a flat prior, following S. Weinberg's argument that there is no compelling reason to prefer a small positive value of $\Lambda$ to $\Lambda=0$, therefore the distribution of $\Lambda$ in the Multiverse should look approximately flat close to $\Lambda=0$. [~\cite{Weinberg_1987}] Secondly, one needs to connect the generation of observers to the star formation history. I will make the minimal assumption that the probability density of generating observers in a universe with cosmological constant $\Lambda$ is simply proportional to the stellar efficiency shown in Fig.~\ref{fig:eff}.

\begin{figure}
    \centering
    \includegraphics[width=0.5\textwidth]{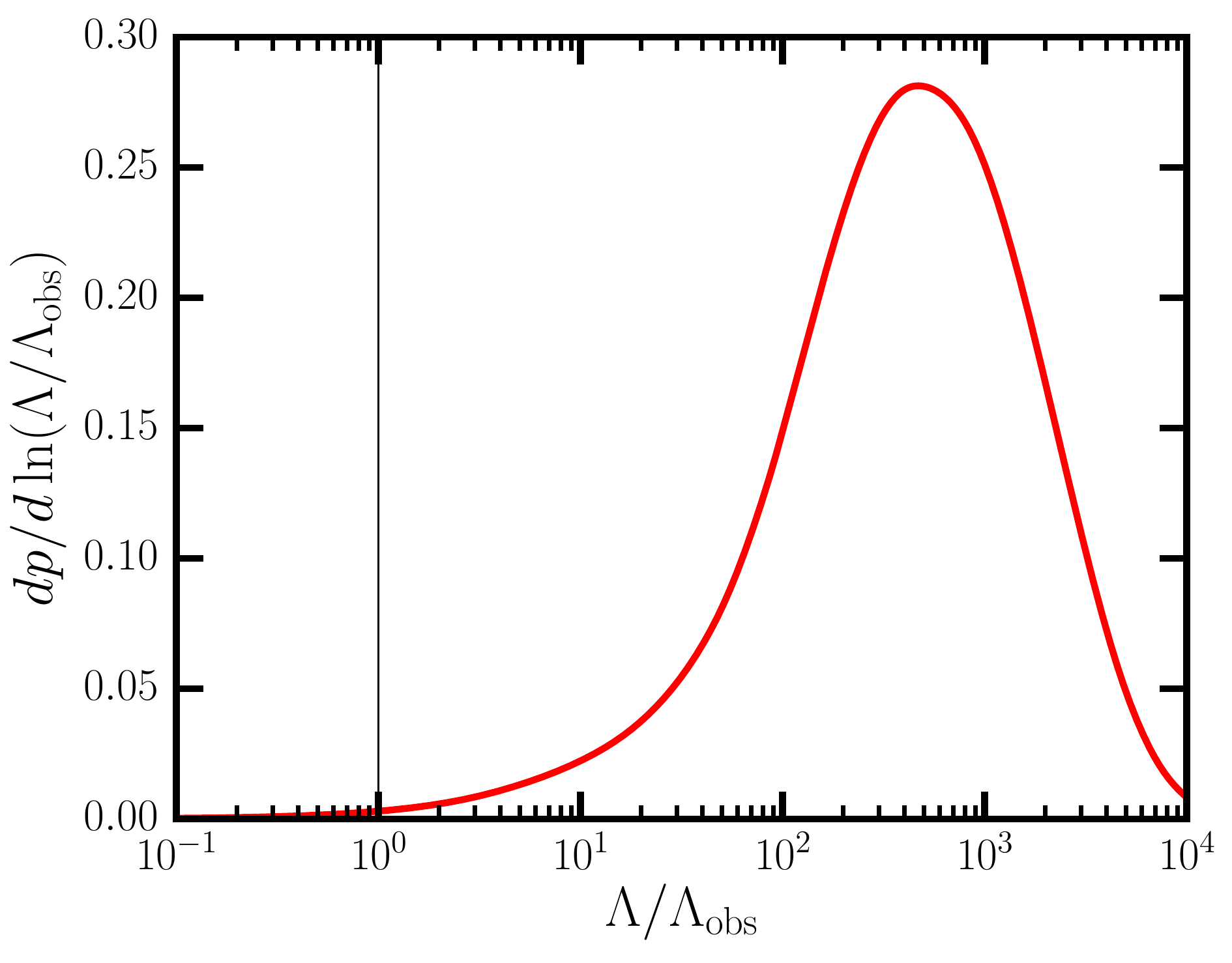}
    \caption{Probability of generating observers vs. $\Lambda$, for a flat \LCDM\ universe, assuming that it is proportional to the global stellar efficiency (Fig.~\ref{fig:eff}), and imposing a flat prior on $\Lambda$. Values $\Lambda \leq \Lambda_{\rm obs}$ are disfavoured.}
    \label{fig:prob}
\end{figure}

The resulting probability distribution is shown in Fig.~\ref{fig:prob}. The peak occurs around $\sim 400-500 \, \Lambda_{\rm obs}$ and not around $\Lambda_{\rm obs}$ as is the case for the stellar efficiency. This is a consequence of the flat prior on $\Lambda$, which gives more weight to larger values of $\Lambda$. This would not be the case if one assumed a different prior, such as a flat prior in $\log_{10}(\Lambda/\Lambda_{\rm obs})$. However, this choice would not allow for $\Lambda<0$. Negative values of $\Lambda$ are possible within the Multiverse ensemble predicted by the landscape of string theory, and I plan to extend my analysis to $\Lambda<0$ in future work.

The preliminary results presented in this manuscript suggest that although $\Lambda_{\rm obs}$ maximises the stellar efficiency, observers' generation actually favours $\Lambda \approx 400-500 \, \Lambda_{\rm obs}$. Instead, $\Lambda_{\rm obs}$ is not anthropically favoured, the probability that $\Lambda \leq \Lambda_{\rm obs}$ being $\sim 0.3\%$. This is even smaller than the $\sim 2\% $ probability found by Barnes et al. (2018), [~\cite{Barnes_2018}] adopting a mass-weighted probability measure. Nevertheless, the conclusion of this manuscript does not rule out anthropic arguments as a viable explanation for $\Lambda_{\rm obs}$. Rather, this work highlights that maximum stellar efficiency does not necessarily imply maximum probability of observers' generation. When invoking anthropic reasoning, it is thus important to bear in mind the subtleties involved with the assumptions on the link between observers' generation and star formation, as well as on the prior on $\Lambda$. 

\section*{Acknowledgments}

This work is done in collaboration with John A. Peacock and Lucas Lombriser. I acknowledge support from the European Research Council, under grant no. 670193, and the STFC consolidated grant no. RA5496.

\section*{References}

\bibliography{Sorini} 

\end{document}